# Lead Sulphide Nanocrystal: Conducting Polymer Solar Cells

*Andrew A. R. Watt\*, David Blake, Jamie H. Warner, Elizabeth A. Thomsen, Eric L. Tavenner, Halina Rubinsztein-Dunlop & Paul Meredith.*

Soft Condensed Matter Physics Group and Centre for Biophotonics and Laser Science, School of Physical Sciences, University of Queensland, Brisbane, QLD 4072 Australia.

\* E-mail: watt@physics.uq.edu.au, Fax: +61 7 3365 1242, Tel: +61 7 3365 1245.

## ABSTRACT

In this paper we report photovoltaic devices fabricated from PbS nanocrystals and the conducting polymer poly (2-methoxy-5-(2'-ethyl-hexyloxy)-p-phenylene vinylene) (MEH-PPV). This composite material was produced via a new single-pot synthesis which solves many of the issues associated with existing methods. Our devices have white light power conversion efficiencies under AM1.5 illumination of 0.7% and single wavelength conversion efficiencies of 1.1%. Additionally, they exhibit remarkably good ideality factors ($n$=1.15). Our measurements show that these composites have significant potential as soft optoelectronic materials.

## I. INTRODUCTION

The first semiconductor nanocrystal:conducting polymer photovoltaic device was reported by Greenham *et al.* in 1996 [1]. Since then, several groups have effectively demonstrated such devices with power conversion efficiencies under AM1.5 light of up to 1.8%. [2,3,4]. Conducting polymers such as MEH-PPV have high hole mobility and low electron mobility [5]. Photovoltaic devices are limited by the minority carrier mobility. Hence, the intrinsic carrier mobility imbalance in MEH-PPV severely limits the performance of pure polymer based photovoltaics. To overcome this imbalance, a second material is often incorporated to act as an electron acceptor and pathway for electron transport. So far, the best devices have been made by incorporating C60 derivatives [6] and cadmium selenide semiconductor nanocrystals [7] into the polymer.

In the case of semiconductor nanocrystals, efficiency improvements have been mainly attributed to altering nanocrystal morphology [3,7]. In these reports, nanocrystals were synthesized separately and subsequently mixed with a conducting polymer. This approach has two shortcomings: firstly the surfactant used to prepare the nanocrystals has to be removed. However, in the majority of cases, a small proportion of this surfactant is incorporated into the final composite and this inhibits efficient charge transfer between nanocrystal and conducting polymer. Secondly, the mixing approach requires the use of co-solvents which adversely effects nanocrystal solubility and polymer chain orientation. Recently, we have developed a new nanocrystal synthesis which eliminates these synthetic problems by using the conducting polymer to control nanocrystal growth [8].

We have chosen lead sulphide (PbS) as our nanocrystal material because, in the quantum regime it has tunable broad band absorption [9,10], electrons and holes are equally confined [11] and excited states are long lived [10]. The electron affinity ($\chi$) of bulk PbS is $\chi$=3.3eV is larger than C60 ($\chi$=2.6eV), this increases the probability of charge separation. Finally lead selenide nanocrystals, which are very similar to PbS, have been shown to possess high efficiency carrier multiplication, this has the potential to exceed the theoretical maximum thermodynamic conversion efficiency [12]. Together, these properties make PbS nanocrystals a material with great potential in polymer-based photovoltaic devices. This paper presents our first device results from this new synthesis.

## II. EXPERIMENTAL

**A. Preparation of Nanocrystal: Conducting Polymer Composite**

The nanocrystal: conducting polymer composite was prepared via a method similar to reference 8. The entire reaction took place in a nitrogen dry-box as follows: A sulphur precursor solution was made by dissolving 0.1g of sulphur flakes in 5ml of toluene. In a typical synthesis, 9ml of toluene, 0.01g of MEH-PPV, 3ml of di-methylsulfoxide DMSO and 0.1g of lead acetate were dissolved in a 20 ml vial on a stirrer-hotplate. All materials where purchased from Sigma Aldrich and used without further purification. With the solution at 160 ºC, 1ml of the sulphur precursor was injected. 0.2ml aliquots where taken every three minutes and injected into 2ml of toluene at ambient temperature. The reaction took approximately 15 minutes to reach completion upon which a black/brown solution resulted. The product was cleaned to remove excess lead and sulphur ions, DMSO and low molecular weight MEH-PPV by adding anhydrous methanol to cause precipitation of the composite material. The sample was centrifuged and the supernatant removed. The precipitate was then redissolved in chlorobenzene and shaken vigourously for 1 hour. The final weight percentage of nanocrystals was then determined gravimetrically. Typically we found 50-60% nanocrystals by weight.

**B. Nanocrystal: Conducting Polymer Composite Material Characterisation**

Transmission electron microscopy (TEM) of the composite material was carried out using a Tecnai 20 Microscope operating at 200kV. Samples where prepared by taking the cleaned product, diluting it and placing a drop on an ultra thin carbon coated copper grid (Ted Pella) with the Formvar removed. The microscope was operated in scanning transmission mode when investigating nanocrystal ensembles and dark field mode for crystal structure.

A Perkin-Elmer λ40 UV-Visible Spectrophotometer was used to obtain absorption spectra of the nanocrystal: conducting polymer solutions in toluene. Measurements were taken in a 10mm quartz cuvette, at a scan speed of 120 nm per minute, and using a band pass of 1nm. The spectral response was adjusted using a toluene reference.

Time-resolved photoluminescence decay measurements were performed using a time-correlated single photon counting spectrometer (Picoquant FluoTime 200). For excitation a Ti:Saphire femtosecond laser was used (Spectra Physics Tsunami), operating with an 80MHz pulse train with individual pulses having a duration of 70 fs at a wavelength of 800nm. This was then frequency doubled using a non-linear BBO crystal for single photon excitation. Excitation intensities were kept constant at below 10 nJ cm$^{-2}$ to minimize photo-oxidization effects and reduce the probability of exciton-exciton annihilation [13]. Thin film samples where prepared by spin casting the nanocrystal: polymer composite material on to clean 25 mm$^2$ glass slides at 1500 rpm for 60 seconds in a nitrogen dry box. Decay characteristics where analysed using FluoFit fluorescence lifetime analysis software (Picoquant).

**C. Photovoltaic Device Fabrication**

25 mm$^2$ ITO substrates purchased from Delta Technologies ($R_s$=4-8Ω) where cleaned thoroughly by ultasoncating for 20 minutes each in the order, Alconox: H$_2$O, H$_2$O, acetone, isopropanol. The substrates where then treated with an Oxygen plasma at a pressure of 200 millitorr, at 40 mA and a frequency of 13.6 MHz for 6 minutes. A blend of poly(3,4-ethylene dioxythiophene) with poly-(styrene sulfonate) (PEDOT/PSS) (HC Starck, Baytron P VP CH 8000) was filtered through a 0.45μm filter and spun cast twice at 2000rpm for 40 seconds onto the oxygen plasma-treated indium-tin oxide (ITO) substrate to act as the device anode. This was then baked in a nitrogen dry-box at 150 °C for 1 hour. The nanocrystal: polymer composite material was then spun cast onto the PEDOT/PSS surface at 1500 rpm for 60 seconds. The films were left to dry for 30 min before aluminum cathodes were deposited by thermal evaporation at a vacuum better than $10^{-5}$ mbar. A five minute anneal was carried out at 80 °C in the nitrogen dry-box. The active device area was 0.04cm$^2$, and the device structure is shown in figure 1.

**D. Photovoltaic Device Characterization**

*1. Experimental Setup*

The electrical properties of devices were measured under flowing argon in an electrically shielded box. Current-voltage characteristics where obtained with a Kiethley 2400 source measurement unit. Simulated solar illumination under AM1.5 global conditions was provided by an Oriel 50W Xenon Arc Lamp with AM1.5 filters at an intensity of 5mW cm$^{-2}$ and single wavelength excitation at 560nm through a monochromator at an intensity of 0.01mW cm$^{-2}$. Quantum efficiencies were measured using a Kiethley 6435 picoammeter, Spex monochromator and Oriel 50W Xenon Arc Lamp. All incident light intensities were measured using a NIST calibrated silicon detector.

*2. Calculating Parasitic Resistances*

In real photovoltaic devices power is dissipated through the resistance of the contacts and leakages at the edge of devices. To account for these effects the equivalent circuit shown in figure 2 is often used [14,15]. For steady state measurements, capacitive effects have been neglected as we see no variation in current with voltage scan rate.

In the high field regime, the series resistance ($R_s$) dominates and can be determined from:

$$R_s = \lim_{V_{out} \to \infty} \left( \frac{dV_{out}}{dI_{out}} \right) \quad (1)$$

In the low field regime, the shunt resistance ($R_{sh}$) dominates and can be determined from:

$$R_{sh} \approx \frac{dV_{out}}{dI_{out}}(V_{out} = 0) \qquad R_s << R_{sh} \quad (2)$$

The current voltage characteristics are largely dependent on the series and shunt resistances. An ideal cell would have a series resistance approaching zero and shunt resistance approaching infinity. A low series resistance means that high currents will flow through the cell at low applied voltages and is due to contact resistance and bulk resistance of the photoactive material. A large shunt resistance results if there are shorts or leaks of photocurrent in the device.

*3. Shockley Model Fitting*

The current voltage characteristics of this type of device can be described by the Shockley equation [14,15,16]

$$I(V) = I_0 \left[ e^{(qV/nkT)} - 1 \right] \quad (3)$$

Where $I_0$ is the saturation current, $q$ the magnitude of the electronic charge, $V$ the applied voltage, $n$ the ideality factor, $k$ Boltzmann's constant and $T$ the absolute temperature. The ideality factor $n$ takes into account recombination and tunneling processes and lies between 1 for an ideal diode and 2 when recombination dominates.

Replacing $V$ by $(V_{out} - I_{out}R_s)$ and including series and shunt resistance equation 3 becomes:

$$I(V) = \frac{R_{sh}}{R_s + R_{sh}}\left(I_0\left[e^{(q(V_{out} - I_{out}R)/nkT)} - 1\right] + \frac{V_{out}}{R_{sh}}\right) \qquad (4)$$

Hence, we obtain an equation to describe the current voltage characteristics for the type of device presented in this paper.

## IV. RESULTS AND DISCUSSION

### A. Material Characterization

#### 1. Transmision Electron Microscopy

We performed transmission electron microscopy (TEM) in order to gain information concerning the nature of nanocrystal growth and quality. Figure 3 shows a dark field scanning TEM image of a nanocrystal ensemble cast on to an ultra-thin amorphous carbon film from a dilute solution. The nanocrystals form within the conducting polymer and they are non-aggregated with an average size of 4nm (±2nm). Figure 4 shows the crystal lattice of an individual nanocrystal and demonstrates that they possess a high degree of crystallinity.

## 2. Spectroscopy

MEH-PPV has an absorption edge at around 560nm corresponding to the lowest energy $\pi$ to $\pi^*$ transition [16]. Figure 5 shows how the absorption changes as PbS nanocrystals assemble as the reaction proceeds. The inclusion of nanocrystals results in an extension of the absorption into the near IR. Using a form of the four-band envelope function formalism proposed by Kang and Wise [18] a theoretical prediction of nanocrystal size can be made for the nanocrystal lowest energy transition. From this we find good agreement between the measured absorption and lowest energy transition derived from TEM particle size [19]. This confirms that our size estimates from TEM are typical for the bulk average.

Fluorescence lifetime measurements can be used to demonstrate electronic coupling between components in such systems. We find that the fluorescence lifetime of the MEH-PPV emission decreases in the composite material as shown in figure 6. We have also observed [8] steady state photoluminescence quenching and no change in photoluminescence spectral shape indicating that this effect is not due to variations in polymer conformation. The MEH-PPV lifetime that we observe for our thin films is consistent with literature values [20]. Analysis of the photoluminescence lifetime decays is presented in Table 1. Both MEH-PPV and composite materials are dominated by a process faster than our instrument resolution ($\tau<40$ps) [21]. Since the magnitude of this component is similar in both samples, and assuming it to be associated with an intrachain process [21] that is unaffected by the nanocrystals, then this allows us to compare the the two slower transitions. In both $\tau_2$ and $\tau_3$ the lifetime is lengthened and amplitude decreased by about 50% in the composite material. Together these results strongly suggest that longer lived MEH-PPV excited states are quenched by the nanocrystal and the two materials are therefore electronically coupled. The rest of the paper examines how this converts to an overall improvement in the materials electrical properties.

## B. Device Characterisation

### 1. Device Efficiency

The current densities as a function of voltage for devices made from MEH-PPV and the composite material in the dark and under illumination are shown in figure 7. It is clear that the inclusion of nanocrystals in the MEH-PPV alters and improves the open circuit voltage ($V_{oc}$) and short circuit current ($I_{sc}$). Device results and experimental conditions are summarized in table 2. The composite cell displays a relatively modest fill factor (0.28-0.30), but respectable power conversion efficiency under white light and single wavelength illumination (0.7 and 1.1% respectively). These results demonstrate that nanocrystals grown using the new synthesis and with equal xyz geometry can be utilized to make a device with efficiency comparable to those made from semiconductor nanocrystal rods or tetrapods [1,2,3,4,7]. It is important to note that the preparation of nanocrystals in a separate surfactant and the subsequent transfer to conducting polymer is not ideal, and device optimization using our "surfactant-free" synthesis may ultimately yield higher efficiencies.

### 2. Parasitic Resistances

From equation (1) and (2) the cell measured in Figure 8 has $R_s=4x10^9 \Omega$ and $R_{sh}=5x10^{11} \Omega$. The large series resistance explains the poor fill factor and low current flow in the device and is most likely a reflection of the quality of the electrical contacting and bulk resistivity. The PEDOT used in construction is not ideal for this application as it has a larger intrinsic resistance than standard Baytron P [22]. The shunt resistance is large indicating that shorts or leakages of photocurrent are minimal in the device.

## 3. Shockley Model

In the device examined we assume that the aluminium contact is ohmic and the ITO:PEDOT contact is a non-ohmic Schottky type contact. Equation 4 was fitted to the measured results using a least squares fit, and the results are shown in figure 7. The fit is good even at high electric fields, the ideality factor is $n$=1.15, which is much better than previous reports for other nanocrystal conducting polymer devices [14]. This means that space charge limited effects are minimal [23]. Conventionally, this would also indicate that loss mechanisms such as recombination due to traps or mid-gap states are small. However there is some ambiguity in this interpreatation since recently Bakueva et al. showed that a PbS nanocrystal: conducting polymer composite material displays negative capacitive effects [24]. This would alter the composite material's charge profile potentially masking these loss mechanisms.

## 5. Current voltage characteristics under illumination

Photocurrent ($I_L$) calibrated for the incident light intensity is expressed by the external quantum efficiency (EQE):

$$EQE(V) = \frac{(I_L - I)V}{P_s} \qquad (5)$$

Where $I$ is the dark current and $P_s$ is the incident light power density

In conventional inorganic solar cells the photocurrent is independent of applied bias, ie EQE is constant as a function of $V$, and $I_L$ can simply be subtracted from the Shockley equation [16]. However, if we plot the external quantum efficiency EQE versus applied bias for MEH-PPV and composite we find that the photocurrent is dependent on the applied bias as shown in figure 8. Therefore the standard Shockley equation under illumination is not valid for these systems. In the MEH-PPV device under reverse bias when the field is in the direction of the built-in electric field, there is increased EQE compared to reverse bias conditions. The composite material exhibits the same behaviour under reverse bias however under forward bias (ie opposing the in built electric field) the EQE increases dramatically. This is a surprising result, a possible explanation is at higher fields the probability of high energy photons absorbed in the nanocrystal relaxing to the band edge is increased [12].

A logarithmic-linear plot of EQE versus Voltage (figure 10) enables us to interrogate the behavior at low bias voltages more clearly. At $V=0V$ and $V=1V$ the EQE in both devices is a minimum. At V=0V the device is operating under its inbuilt electric field arising from the difference in work functions between ITO:PEDOT and Aluminium contacts. $V=1V$ corresponds to when the applied field opposes the inbuilt electric field which results in a minimum EQE. This tells us that the built-in electric field profile at the material contacts is poor and there are lots of shallow charge trapping sites. The minimum at $V=0.6V$ in the MEH-PPV device is a result of the carrier imbalance in MEH-PPV, whereby only excitons within one diffusion length of the material junction can be separated due to space charge effects and poor conductivity. This corresponds to the open circuit voltage and the residual dark current minima [25].

## 6. Incident photon conversion efficiency

Figure 11 presents the incident photon conversion efficiency (IPCE) which reaches a maximum of 21% at 500nm. Disappointingly the nanocrystal absorption from the aliquots in figure 3 is not reflected in the ICPE or solid film absorption. A possible explanation for this is that the nanocrystals quantum confinement properties change when a percolated network is formed in the solid state.

## V. CONCLUSIONS

In conclusion, we have demonstrated a plastic solar cell made from a PbS:MEH-PPV composite. Most significantly, we have shown that our new surfactant-free nanocrystal synthesis is effective and that nanocrystals with equal xyz geometry can be utilized to make respectable photovoltaic devices. Further efficiency improvements of these devices could be gained by tuning the nanocrystal concentration and morphology, and optimizing the active layer thickness. We find that devices adhere to the Shockley equation in the dark but not under light conditions. From a study of the dark current we demonstrated devices with remarkably good ideality factors ($n$=1.15). We have also shown that external quantum efficiency is strongly dependent on applied bias. Overall our measurements show that these composites have significant potential as soft optoelectronic materials. Further work is underway to understand the exact nature of carrier transport mechanisms involved.

## ACKNOWLEDGMENTS

The work was funded by the Australian Research Council and through the University of Queensland Research Infrastructure Scheme. TEM was performed at the University of Queensland Centre for Microscopy and Microanalysis, AARW thanks the University of Queensland for an International Postgraduate Research Scholarship.

# References

[1]     Greenham N C, Peng X, Alivisatos A P 1996 *Phys. Rev. B* **54** 17628

[2]     Liu J, Tanaka T, Sivula K, Alivisatos A P, Frechet J M J 2004 *J. Am. Chem. Soc.* **126** 6551

[3]     Sun B, Marx E, Greenham N C 2003 *Nano Letters* **3** 961

[4]     Arici A, Sariciftci N S, Meissner D 2003 *Adv. Funct. Mater.* **13** 165

[5]     Antoniadus H, Abkowitz M A, Hsieh B R 1994 *Appl. Phys. Lett.* **65** 2030

[6]     Shaheen S E, Brabec C J, Sarcifitci N S, Padinger F, Fromherz T, Hummelen J C 2001 *Appl. Phys. Lett.* **78** 841

[7]     Huynh W U, Dittmer J J, Alivisatos A P 2002 *Science* **295** 2425

[8]     Watt A, Thomsen E, Meredith P, Rubinsztein-Dunlop H 2004 *Chem. Commun.* **20** 2334

[9]     Hines M A, Scholes G D 2003 *Adv. Mater.* **15** 1844

[10]    Warner J H, Thomsen E, Watt A R, Heckenberg N R, Rubinsztein-Dunlop H *Nanotechnology* DOI10.1088/0957

[11]    Wise F W 2000 *Accounts Chem. Res.* 2000 **33** 773

[12]    Schaller R D, Klimov V I 2004 *Phys. Rev. Lett.* **92** 186601

[13]    Samuel I, Rumbles G, Collison C, Friend R, Moratti S, Holmes A. 1997 *Synth. Met.* **84** 497

[14]    Huynh W U, Dittmer J J, Teclemariam N, Milliron D J, Alivisatos A P, Barnham K W J 2003 *Phys. Rev. B* **11** 5326

[15]    Nelson J, 2003 The Physics of Solar Cells (Imperial College Press, London)

[16]    Shockley W, 1961 *J. Appl. Phys.* **32** 510

[17]    Martin S J, Bradley D D C, Lane P A, Mellor H, Burn P L 1999 *Phys. Rev. B* **59** 15133

[18]    Inuk K, Wise F W, 1997 *J. Opt. Soc. Am. B* **14** 1632

[19]    Watt A A R, Rubinsztein-Dunlop H, Meredith P *submitted J. Am. Chem. Soc.*

[20]     Samuel I D W, Crystall B, Rumbles G, Burn P L, Holmes A B, Friend R H, 1993 *Chem. Phys. Lett.* **213** 472

[21]     Hendry E, Schins J M, Candeias L P, Siebbeles L D A, Bonn M 2004 *Phys. Rev. Lett.* **92 19**6601

[22]      http://www.hcstarck-echemicals.com/request/pages/baytron/p_oleds_generalinfo.html

[23]     Tagmouti S, Outzourhit A, Oueriagli A, Khaidar M, Elyacoubi M, Evrard R, Ameziane E L 2000 *Thin Solid Films* **379** 272

[24]     Bakueva L, Konstantatos G, Musikhin S, Ruda H E Shik A 2004 *Appl. Phys. Lett.* **12** 3567

[25]     Breeze A J, Schlesinger Z, Carter S A, Brock P J 2001 *Phys. Rev. B* **64** 125205

[26]     Salaneck W R 1997 *Phil. Trans.Mathematical, Physical and Engineering Sciences* **355** 789

# Figures and Captions

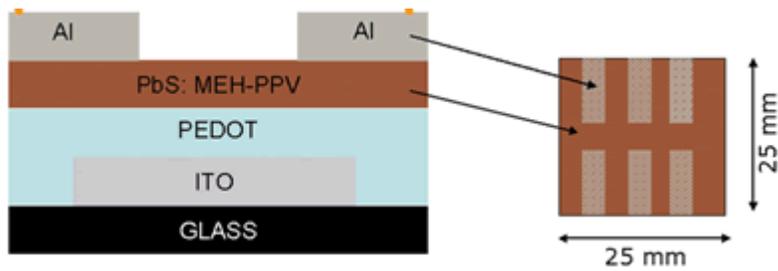

**Figure 1.** Photovoltaic device structure.

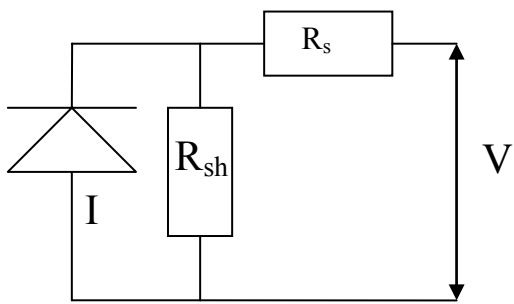

**Figure 2.** Eqivalent circuit diagram for a photovoltaic device in the dark including series ($R_s$) and shunt ($R_{sh}$) resistances.

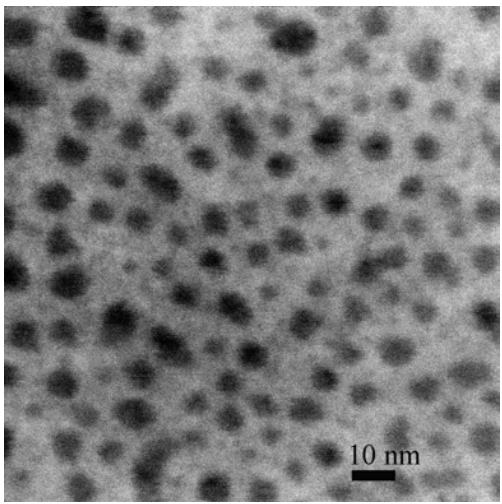

**Figure 3.** Dark field scanning transmission electron microscopy image of a dilute sample of PbS nanocrystals.

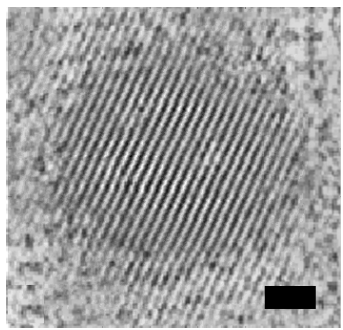

**Figure 4.** High resolution dark field TEM image of the lattice planes in a single nanocrystal (bar=1nm).

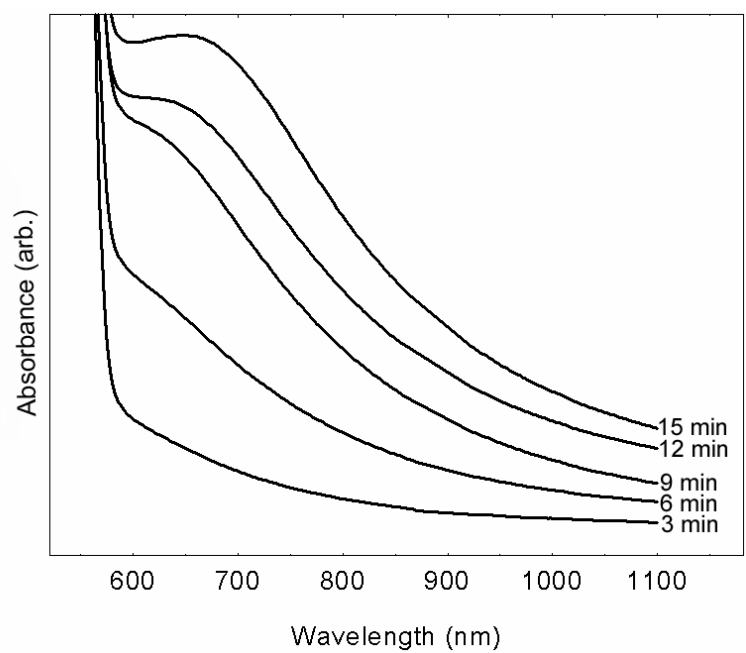

**Figure 5.** Change in absorption as aliquots are taken as the reaction proceeds.

| MEH-PPV ($\chi^2$=1.1) | | Composite ($\chi^2$=1.6) | |
|---|---|---|---|
| Lifetime | Amplitude | Lifetime | Amplitude |
| $\tau_1$<40ps | 552049 | $\tau_1$<40ps | 568114 |
| $\tau_2$=0.16ns | 65528 | $\tau_2$=0.2ns | 30686 |
| $\tau_3$=0.7ns | 2635 | $\tau_3$=1.1ns | 1536 |

**Table 1.** Summary of fluorescent lifetime components with amplitudes.

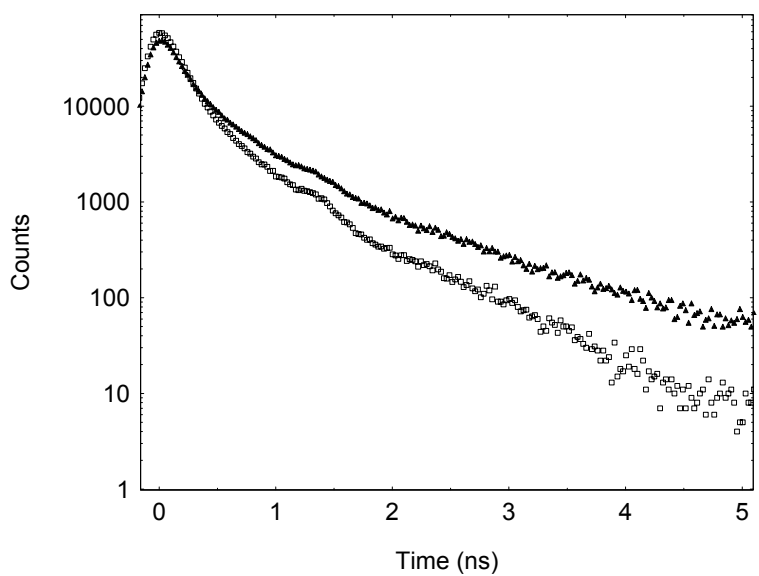

**Figure 6.** MEH-PPV Fluorescent lifetime (excitation 410nm, emission 600nm) of pure MEH-PPV (filled triangles) and MEH-PPV: PbS composite (empty squares).

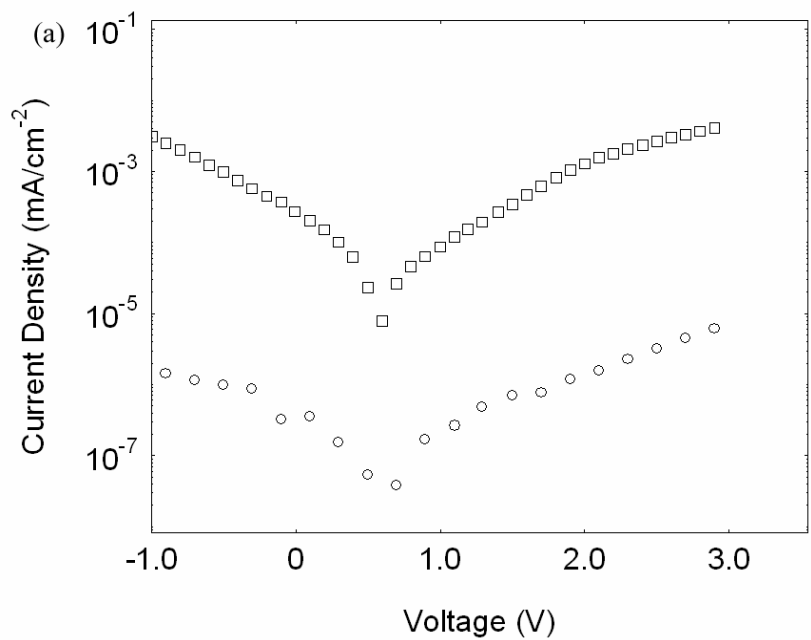

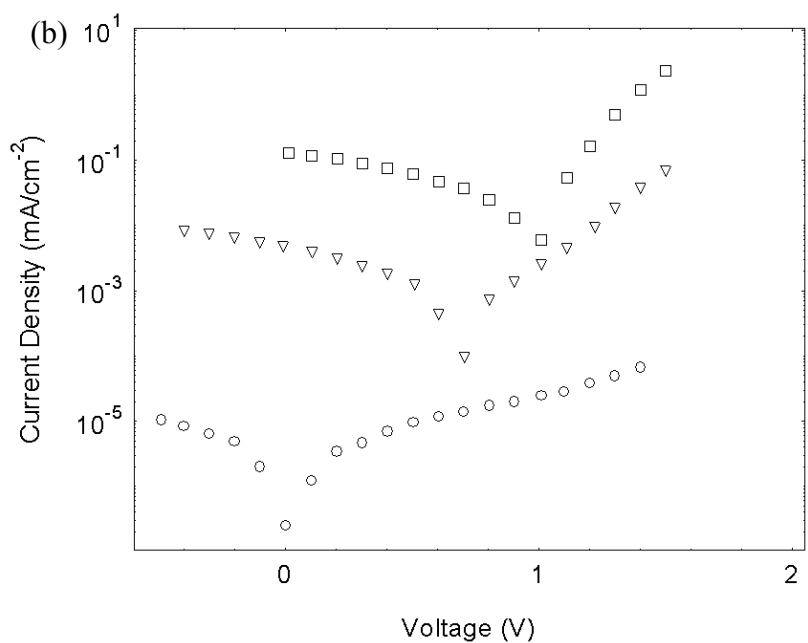

**Figure 7.** (a) Current density vs voltage for a MEH-PPV solar cell in the dark (circles) and simulated AM1.5 global light at an intensity of 5 mW cm$^{-1}$ (squares). (b) Current density vs voltage for a PbS: MEH-PPV solar cell in the dark (circles), 0.01 mW cm$^{-1}$ illumination at 560nm (triangles) and simulated AM1.5 global light at an intensity of 5 mW cm$^{-1}$ (squares).

| Wavelength | 560nm | AM1.5 | AM1.5 |
|---|---|---|---|
| Material | Composite | Composite | MEH-PPV |
| Incident Power (mW/cm$^2$) | 0.01 | 5 | 5 |
| $V_{oc}$ (V) | 0.7 | 1 | 0.6 |
| $I_{sc}$ (mA/cm$^2$) | -0.005 | -0.13 | -5.5x10$^{-6}$ |
| Fill Factor | 0.30 | 0.28 | 0.2 |
| Power Conversion Efficiency (%) | 1.1 | 0.7 | 0.0003 |

**Table 2.** Summary of solar cell device characteristics.

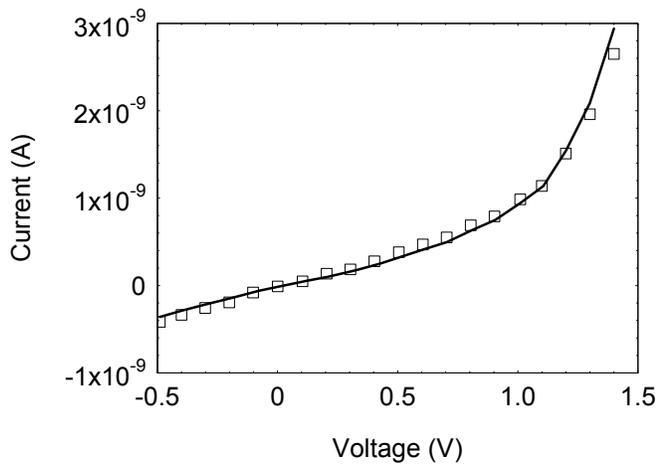

Figure **8.** Experimental dark current voltage (circles) and Schockley equation fit (line).

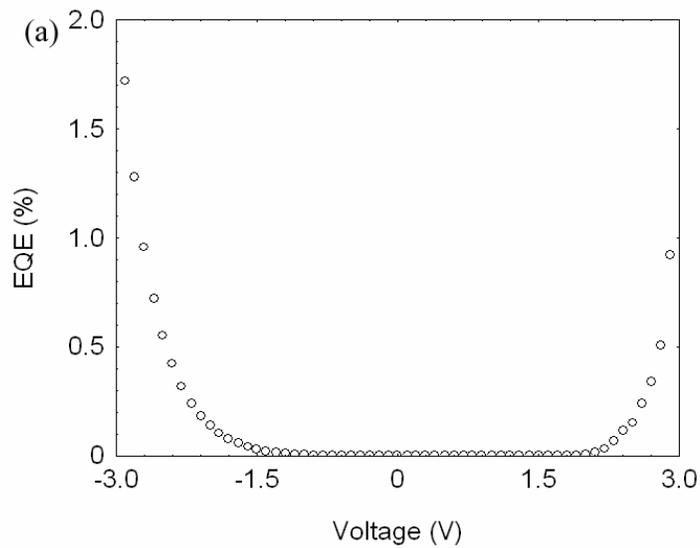

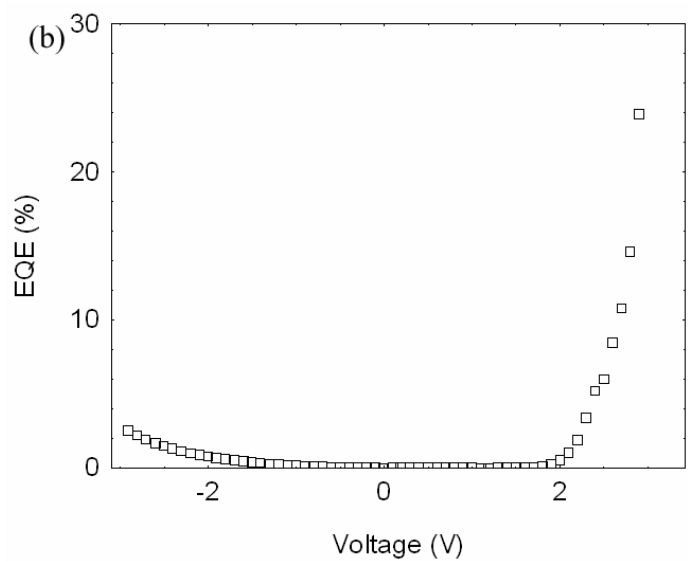

Figure **9.** EQE versus bias voltage for (a) MEH-PPV and (b) composite under simulated AM1.5 global light at an intensity of 5 mW cm$^{-1}$.

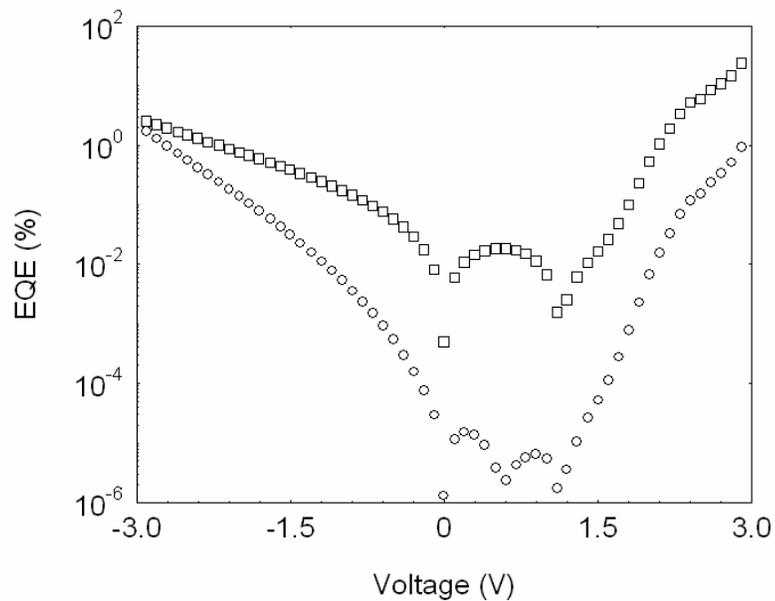

**Figure 10.** Logarithmic plot of EQE versus bias voltage for MEH-PPV (circles) and Composite (squares).

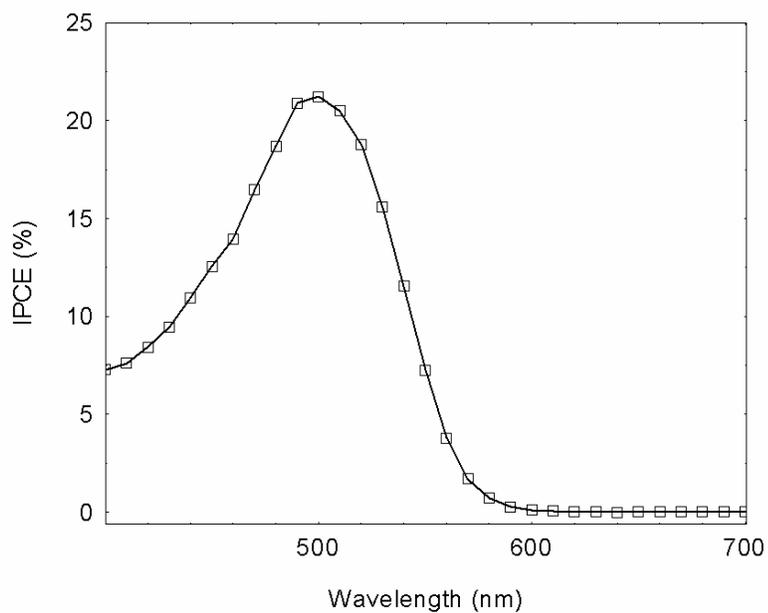

**Figure 11.** Incident photon conversion efficiency (IPCE) for a PbS: MEH-PPV solar cell.

21